\documentclass[%
 notitlepage,
 reprint,
 superscriptaddress,
 amsmath,amssymb,
 aps,
]{revtex4-1}

\usepackage{lipsum}
\usepackage{graphicx}
\usepackage{dcolumn}
\usepackage{bm}

\usepackage{epstopdf}

\newcommand{\bra}[1]{\langle #1 \vert}
\newcommand{\ket}[1]{\vert #1 \rangle}

\begin{document}

\preprint{APS/123-QED}

\title{High-fidelity state transfer through long-range correlated disordered quantum channels}

\author{Guilherme M. A. Almeida}
\email{gmaalmeida.phys@gmail.com}
\affiliation{%
 Instituto de F\'{i}sica, Universidade Federal de Alagoas, 57072-900 Macei\'{o}, AL, Brazil
}%
\author{Francisco A. B. F. de Moura}
\affiliation{%
 Instituto de F\'{i}sica, Universidade Federal de Alagoas, 57072-900 Macei\'{o}, AL, Brazil
}%
\author{Marcelo L. Lyra}
\affiliation{%
 Instituto de F\'{i}sica, Universidade Federal de Alagoas, 57072-900 Macei\'{o}, AL, Brazil
}%

\date{\today}

\begin{abstract}
We study quantum-state transfer in $XX$ spin-$1/2$ chains where 
both communicating spins are weakly coupled to a channel featuring disordered on-site magnetic fields.
Fluctuations are modelled by long-range correlated sequences
with self-similar profile obeying a power-law spectrum. 
We show that the channel is able to perform an almost perfect 
quantum-state transfer in most of the samples
even in the presence of significant amounts of disorder 
provided
the degree of those correlations is strong enough.
In that case, we also show that the lack of mirror symmetry 
does not affect much the likelihood of having high-quality outcomes.
Our results advance a further step 
in designing robust devices for   
quantum communication protocols. 

\end{abstract}

\maketitle


\section{\label{sec1}Introduction}


Transmitting quantum states and establishing entanglement between
distant parties (say Alice and Bob)
are crucial tasks in quantum information processing protocols \cite{cirac97, kimble08}.
%
In this direction, 
spin chains have been widely addressed as quantum channels 
for (especially short-distance) communication protocols since 
proposed in Ref. 
\cite{bose03} that
spin chains can be  
used for carrying out transfer of quantum information with minimal control, i.e.,  
no manipulation is required during the transmission. 
Basically, Alice prepares and
sends out an arbitrary qubit state through the channel and Bob only needs 
to make a measurement at some prescribed time. The evolution itself is given
by the natural dynamics of the system. 

Since then, several schemes for high-fidelity quantum-state transfer (QST) 
\cite{bose03,christandl04,plenio04,osborne04,wojcik05,*wojcik07,li05,huo08, gualdi08, banchi10, *banchi11,*apollaro12,lorenzo13,lorenzo15, almeida16} 
and entanglement creation and distribution 
\cite{amico04, plenio05, apollaro06,*plastina07,*apollaro08,venuti07, cubitt08, giampaolo09, *giampaolo10, gualdi11, estarellas17, *estarellas17scirep, almeida17-1} 
in spin chains have been put forward. 
For instance, 
it was discovered that \textit{perfect} QST can be achieved in mirror-symmetric chains by a judicious tuning
of the spin-exchange couplings over the entire chain 
\cite{christandl04, plenio04} (see \cite{feder06} for a generalization). 
While this scheme allows one to perform QST with unit fidelity
for arbitrarily-large distances, it is not an easy task, on the practical side, 
to engineer the whole chain with the desired 
precision, what makes this configuration very sensitive to
perturbations \cite{dechiara05, zwick11, *zwick12}.  
An alternative less-demanding approach is based on optimizing the outermost
couplings of a uniform channel so that the linear part of the spectrum dominates the dynamics \citep{banchi10}. 
One can also encode the information using multiple spins to 
send dispersion-free Gaussian wave-packets through the channel \cite{osborne04}. 
Another class of protocols relies on setting \textit{very weak} couplings 
between the end spins (those being the sender and receiver sites) 
and the bulk of the chain \cite{wojcik05, li05,venuti07,huo08, gualdi08,giampaolo09, giampaolo10, almeida16} 
in order to effectively reduce the operating Hilbert space 
to that of a two- or three-site chain, depending on 
the
resonance conditions. 
That way, it is possible to carry out QST with close-to-unit fidelity. 
A similar strategy is to apply strong magnetic fields at the sender and receiver spins (or on their nearest neighbors) \cite{plastina07, lorenzo13, paganelli13}.    




Each of the aforementioned schemes has its 
own peculiarities 
but there is one detail that can seriously compromise 
the protocol regardless of the engineering scheme being used, that is disorder.
%
Fluctuations either in the local magnetic fields or in the coupling strengths 
are inevitably present either due to manufacturing errors
or dynamical spurious factors 
hence leaving us far from the desired output.
Needless to say, finding out ways to overcome such difficulties and 
and testing the robustness of various schemes against such experimental imperfections are
of great importance and have been done extensively \cite{dechiara05,fitzsimons05,burgarth05,tsomokos07,giampaolo10, petrosyan10,yao11,zwick11, bruderer12,kay16}.
Among many possible configurations to realize high-quality QST, in the presence of disorder it should be much more preferable to choose a channel in which 
the sender and receiver spins do not heavily depend upon. Having that in mind, 
those setups featuring communicating parties weakly coupled to the channel \cite{wojcik05}
seem to be a promising choice \cite{yao11}. A combined approach involving modulated couplings with weakly coupled spins has been also put forward in \cite{bruderer12}. 
Still, the slightest amount of disorder is already capable of promoting 
Anderson localization effects \cite{anderson58, *abrahams79} or, even worse, destroying
the symmetry of the channel \cite{albanese04}.
That is not necessarily true, however, in the case of \textit{correlated} disorder. 
The breakdown of Anderson localization has been reported when short- \cite{flores89, *dunlap90,*phillips91} or long-range correlations \cite{demoura98,izrailev99,kuhl00,lima02, *demoura02,*nunes16,demoura03,adame03, gonzales14, almeida17-1} are present in disordered 1D models. In particular, the latter case
finds a set of extended states in the middle of the band with well detached
mobility edges thereby signalling an Anderson-type metal-insulator transition \cite{demoura98,izrailev99}. This is also manifested in low-dimensional spin chains \cite{lima02, almeida17-1}.

Correlated fluctuations takes place in many stochastic processes in nature (see, e.g., Refs. \cite{lam92,peng92,carreras98, carpena02}) and therefore
shall not be ruled out 
when designing protocols for quantum information processing in solid-state devices \cite{dechiara05,burgarth05}.
Here, we will see that it indeed makes a dramatic difference 
in the performance of QST protocols based on weakly-coupled end spins.  
Specifically, we consider an one-dimensional $XX$ spin chain
in which
the local magnetic fields (on-site potentials) of the channel 
follow a long-range correlated disordered distribution
with power-law spectrum $S(k) \propto 1/k^{\alpha}$, with $k$ being the corresponding wave number and
$\alpha$ being a characteristic exponent governing the degree of such correlations. 
We show that when perturbatively attaching two communicating (end) spins to the channel
and setting their frequency to lie in the middle of the band, 
we are still able to perform nearly perfect QST rounds 
in the presence of correlated disorder.
%
Surprisingly, it happens even
in the presence of considerable amounts of asymmetries in the channel.
The reason for that is the appearance of extended states in the middle of the band
which
offers the necessary end-to-end \textit{effective} symmetry
thereby supporting the occurrence of Rabi-like oscillations between the sender and receiver spins. 
We show that perfect mirror symmetry, despite being very convenient for QST protocols, is not a crucial factor as long as there exists a proper set of delocalized eigenstates in the channel. 


In the following, Sec. \ref{sec2}, we 
introduce the $XX$ spin Hamiltonian with on-site long-range correlated disorder. 
In Sec. \ref{sec3} 
we derive an effective two-site Hamiltonian that accounts for the way both communicating parties are coupled to the channel.  
In Sec. \ref{sec4} we investigate how the channel responds to disorder 
by looking at the resulting effect on the
localization and symmetry properties. 
In Sec. \ref{sec5} we display the results for the QST fidelity and 
our final remarks
are addressed in Sec. \ref{sec6}.

\section{\label{sec2}Spin-chain Hamiltonian}


We consider a 
pair of spins (communicating parties) coupled
to a one-dimensional quantum channel 
consisting altogether of an open spin-$1/2$ chain featuring 
$XX$-type exchange interactions described by 
Hamiltonian 
$\hat{H} = \hat{H}_{\mathrm{ch}}+\hat{H}_{\mathrm{int}}$
with ($\hbar = 1$)
\begin{equation} \label{Hchannel}
\hat{H}_{\mathrm{ch}} = \sum_{i=1}^{N}\dfrac{\omega_{i}}{2}(\hat{1}-\hat{\sigma}_{i}^{z})-\sum_{\langle i,j \rangle}\dfrac{J_{i,j}}{2}(\hat{\sigma}_{i}^{x}\hat{\sigma}_{j}^{x}
+\hat{\sigma}_{i}^{y}\hat{\sigma}_{j}^{y}), 
\end{equation}
where $\hat{\sigma}_{i}^{x,y,z}$ are the Pauli operators for the $i$-th spin, 
$\omega_{i}$ is the local (on-site) magnetic field, 
and $J_{i,j}$ is the exchange coupling strength between 
between nearest-neighbor nodes. 
Supposing the sender ($s$) and receiver ($r$) spins are connected to nodes $1$ and $N$
from the channel at rates $g_{s}$ and $g_{r}$, respectively, the interaction part reads
\begin{align} \label{Hint}
\hat{H}_{\mathrm{int}} &=\dfrac{\omega_{s}}{2}(\hat{1}-\hat{\sigma}_{s}^{z})+\dfrac{\omega_{r}}{2}(\hat{1}-\hat{\sigma}_{r}^{z})\\ \nonumber
& \quad-
\dfrac{g_{s}}{2}(\hat{\sigma}_{s}^{x}\hat{\sigma}_{1}^{x}
+\hat{\sigma}_{s}^{y}\hat{\sigma}_{1}^{y})+
\dfrac{g_{r}}{2}( \hat{\sigma}_{r}^{x}\hat{\sigma}_{N}^{x}
+\hat{\sigma}_{r}^{y}\hat{\sigma}_{N}^{y}).
\end{align}
%
Note that since $\hat{H}$ conserves the total magnetization of the system, i.e., 
$\left[ \hat{H}, \sum_{i} \hat{\sigma}_{i}^{z}\right] =0$, the
Hamiltonian can be split into independent subspaces with fixed
number of excitations. Here we focus on the single-excitation Hilbert space
spanned by states of the form $\ket{i} = \hat{\sigma}_{i}^{+}\ket{\downarrow\downarrow\ldots\downarrow}$ with $i = r, s, 1,\ldots,N$, that means every spin pointing down but the one located at the $i$-th position. 
In this case, we end up with a 
hopping-like matrix with $N+2$ dimensions.
Indeed 
$\hat{H}$ can be mapped onto a system describing non-interacting spinless fermions through
the Jordan-Wigner transformation.

Let us now make a few assumptions in regard to the channel described by Hamiltonian (\ref{Hchannel}). Here we 
consider the spin-exchange coupling strengths to be uniform $J_{i,j}\rightarrow J$ and, 
in order to study the robustness of the channel against disorder we introduce
correlated static fluctuations on the 
on-site magnetic field $\omega_{n}$, $n=1,\ldots,N$.
%
A straighforward way to generate random sequences featuring internal long-range correlations is through 
the trace of the fractional Brownian motion with power-law
spectrum $S(k)\propto1/k^{\alpha}$
\cite{demoura98, adame03}
\begin{equation} \label{disorder}
\omega_{n} = \sum_{k=1}^{N/2}k^{-\alpha/2}
\mathrm{cos}\left( \dfrac{2\pi n k}{N} + \phi_{k} \right),
\end{equation} 
where 
$k=1/\lambda$, is the inverse modulation wavelength, 
$\lbrace \phi_{k} \rbrace$ are random
phases distributed uniformly within $\left[0,2\pi \right]$, 
and $\alpha$ controls the degree of correlations.
This parameter is related to the so-called 
Hurst exponent \cite{fractalsbook},  
$H = (\alpha -1)/2$, which characterizes 
the self-similar character of a given sequence. When
$\alpha= 0$, we recover the case of
uncorrelated disorder (white noise) and for 
$\alpha>0$ underlying long-range correlations take place. The resulting long-range correlated sequence becomes nonstationary for $\alpha > 1$.
Furthermore, according to the usual terminology,
when $\alpha > 2$ ($\alpha < 2$) the series increments become persistent (anti-persistent). 
Interestingly, this brings about serious consequences on the spectrum
profile of the system. As shown in \cite{demoura98, adame03}, when $\alpha>2$ there occurs
the appearance of delocalized states in the middle of the one-particle spectrum band.
In the QST scenario with weakly-coupled spins $r$ and $s$, i.e. $g_{s},g_{r} \ll J$, 
that promotes a strong enhancement in the likelihood 
of disorder realizations with very-high fidelities $F$, most of them yielding $F\approx1$. 
This will be elucidated along the paper.

Hereafter we set the sequence generated by Eq. (\ref{disorder})
to follow a normalized distribution, that is
$\omega_{n} \rightarrow 
\left( \omega_{n}-\langle\omega_{n}\rangle \right) / \sqrt{\langle\omega_{n}^2\rangle-\langle\omega_{n}\rangle^2}$.
We also stress that such a disordered distribution
has no typical length scale which is a
property of many natural stochastic series \cite{bak96}.

\section{\label{sec3}Effective two-site description}

We now work out a perturbative approach to write down a proper representation of an effective Hamiltonian involving only the sender and receiver spins provided they are 
very weakly coupled to the channel. 
Intuitively, we expect they span their 
own subspace with renormalized parameters and thus QST takes place via effective Rabi oscillations between them \cite{wojcik05, gualdi08, lorenzo13, almeida16}. Our goal here is 
to investigate the influence of disorder in such subspaces and see about how 
much asymmetry they are able to tolerate. 

Here, we follow the procedure adopted in Refs. \cite{wojcik07, li05}. To begin with, let us express the channel Hamiltonian, Eq. (\ref{Hchannel}), in
terms of its eigenstates $\lbrace \ket{E_{k}} \rbrace$ with corresponding (nondegenerate) frequencies $\lbrace E_{k} \rbrace$ and recast $\hat{H} = \hat{H}_{0}+\hat{V}$, such that
\begin{align}
\hat{H}_{0} = \omega_{s}\ket{s}\bra{s} +  \omega_{r}\ket{r}\bra{r}  + \sum_{k}E_{k} \ket{E_{k}}\bra{E_{k}}, \\ 
\hat{V} = \epsilon \sum_{k} \left( g_{s}a_{sk} \ket{s}\bra{E_{k}} + g_{r}a_{rk} \ket{r}\bra{E_{k}} + \mathrm{H.c.} \right)
\end{align}
are now the free and perturbation Hamiltonians, respectively, with $\epsilon$
being a perturbation parameter,
$a_{sk}\equiv\langle 1 \vert E_{k} \rangle$, and 
$a_{rk}\equiv\langle N \vert E_{k} \rangle$. Herein we set units such that $J=1$ for convenience.

If we consider that both $\omega_{s}$ and $\omega_{r}$ do not match any of the normal frequencies $E_{k}$
of the channel and set $\epsilon g_{s}$ and $\epsilon g_{r}$ to be very weak so as to not disturb the nearby
modes, we expect reaching an 
effective Hamiltonian of the form $\hat{H}_{\mathrm{eff}} = \hat{H}_{\mathrm{ch}}\oplus\hat{H}_{sr}$
up to some leading order in $\epsilon$, where $\hat{H}_{sr}$ is the decoupled two-spin Hamiltonian which contains all the valuable information on the way the sender and receiver spins ``feel'' the spectrum of the channel. The trick to find $\hat{H}_{\mathrm{eff}}$
is quite straightforward \cite{wojcik07}. Suppose there is a transformation 
$\hat{H}_{\mathrm{eff}} = e^{i\hat{S}}\hat{H}e^{-i\hat{S}}$, with $\hat{S}$ being a Hermitian operator which we properly choose to be of the form
\begin{equation}
\hat{S} = i\epsilon \sum_{k}\left( \dfrac{g_{s}a_{sk}}{E_{k}-\omega_{s}}\ket{s}\bra{E_{k}} + \dfrac{g_{r}a_{rk}}{E_{k}-\omega_{r}}\ket{r}\bra{E_{k}}+\mathrm{H.c.} \right).
\end{equation}
This choice is very convenient 
because it rules out the first order terms 
$\hat{V}+i[\hat{S},\hat{H}_{0}] = 0$
and, up to second-order perturbation theory, we are then left with
\begin{equation}
\hat{H}_{\mathrm{eff}} = \hat{H}_{0}+ i[\hat{S},\hat{V}] +\dfrac{i^{2}}{2!}[\hat{S},[\hat{S},\hat{H_{0}}]] + O(\epsilon^{3}). 
\end{equation}
By inspecting the above equation, we see that spins $r$ and $s$ are now
decoupled from the rest of the chain, as we intended to. 
The corresponding Hamiltonian projected onto $\lbrace\ket{s}, \ket{r} \rbrace$ then reads
\begin{equation}\label{Heff2}
\hat{H}_{sr}=
\begin{pmatrix}
h_{s} & -J'\\ 
-J' & h_{r}
\end{pmatrix},
\end{equation}
with 
\begin{equation}\label{weff}
h_{\nu} = \omega_{\nu}-\epsilon^{2} g_{\nu}^{2} \sum_{k}\dfrac{|a_{\nu k}|^{2}}{E_{k}-\omega_{\nu}},
\end{equation}
$\nu \in \lbrace s, r \rbrace$, and
\begin{equation}\label{Jeff}
J' = \dfrac{\epsilon^{2}g_{s}g_{r}}{2}\sum_{k}\left( \dfrac{a_{sk}a_{rk}}{E_{k}-\omega_{s}} + \dfrac{a_{sk}a_{rk}}{E_{k}-\omega_{r}} \right).
\end{equation}
Note that we are assuming all parameters to be real.  Hamiltonian (\ref{Heff2}) describes a two-level system which performs Rabi-like oscillations in a time scale 
set by the inverse of the gap between its normal frequencies. In order to have as perfect as possible QST one should guarantee that $h_{s} = h_{r}$. This is
automatically fulfilled, given $\omega_{s} = \omega_{r}$ and $g_{s} = g_{r} = g$, for mirror-symmetric chains since $|a_{sk}| = |a_{rk}|$ for every $k$.  
In that case, for a noiseless uniform channel and in the limit of very weak outer couplings, which implies in the validity of Hamiltonian (\ref{Heff2}), 
an initial state prepared in $\ket{s}$ 
will evolve in time to $\ket{r}$ with nearly unit amplitude at times $\tau J = n\pi / (2J')= n\pi / (2\epsilon^{2}g^{2})$, with $n$ being an odd integer \cite{wojcik05, wojcik07}.
Note that as $N$ increases more eigenstates get in the middle of the spectrum and thus 
$\epsilon g_{\nu}$ must be adjusted accordingly (we shall drop out the perturbation parameter $\epsilon$ hereafter).

In summary, in Rabi-type QST protocols \cite{wojcik05,gualdi08,lorenzo13,almeida16}, a pair of eigenstates of the form $\ket{\psi^{\pm}} \approx (\ket{s}\pm \ket{r})/\sqrt{2}$
is ultimately responsible for the fidelity of the transfer.  
We remark that, 
for certain classes of channels, such as uniform or dimerized 
\cite{venuti07, ciccarello11,almeida16}, one
can obtain analytical forms for those states using perturbation theory as well as work
out the corresponding discrete normal frequencies.  
The form expressed by Eq. (\ref{Heff2}), however, 
is general and more suited for our purposes, not to mention we are dealing with disordered channels. 

We also would like to mention that one can induce an effective three-site
system by properly tuning $\omega_{s} = \omega_{r} = E_{k}$ for a given $k$.
In that case, the transfer is directly \textit{mediated} by the corresponding eigenstate \cite{wojcik07, yao11, paganelli13}. 
Likewise, whenever perfect symmetry between sites $1$ and $N$ is available, which 
corresponds to equal off-diagonal rates in the effective $3\times3$ hopping matrix, QST can be similarly performed with nearly perfect fidelity in the limit of very small $g_{\nu}$ \cite{wojcik07}.
We do not deal with this scenario here because in our disordered chain there 
will be no fixed normal frequencies to tune with since
each sample features a different sequence generated by Eq. (\ref{disorder}).

\section{\label{sec4}Disordered channel properties}

While spatial symmetry is an essential ingredient
in the design of quantum communication protocols in spin chains,
there is no guarantee that all chain parameters will come out as planned. 
Experimental imperfections may induce disorder and hence spoil the intended output. 
In 1D tight-binding models, pure (uncorrelated) disorder yields
the so-called phenomenon of Anderson localization \cite{anderson58} 
in which every eigenstate
becomes exponentially localized around a given site, say $x_{0}$,
$\langle x\vert E_{k}\rangle \sim e^{-\frac{|x-x_{0}|}{\xi_{k}}}$, 
where $\xi_{k}$ is the localization length \cite{thouless74}. 
Now let us discuss the consequences of that on the two-site effective Hamiltonian, 
Eq. (\ref{Heff2}). 
Disorder acts on it by inducing a (undesired) detuning $\Delta \equiv h_{s}-h_{r}$. 
At first glance, one could naively 
think of masking this effect by setting
$J'\gg \Delta$ only to realize that all the Hamiltonian 
parameters heavily depend upon the very same factors. First and foremost, 
they are built from the overlap, $a_{sk}$ and $a_{rk}$, between the 
spins they are connected to (the outer spins of an open linear chain) and \textit{each} normal mode $k$ of the channel. The presence of disorder then promotes a tremendous asymmetry 
in the channel at the same time it decreases $J'$, because it turns out to be very unlikely a given eigenstate $\ket{E_{k}}$ will 
simultaneously feature non-negligible
amplitudes 
in $\ket{1}$ and $\ket{N}$ thereby diminishing the contribution of each term of the sum in Eq. (\ref{Jeff}). 
As a consequence, 
the subspace spanned by $\ket{s}$ and $\ket{r}$
becomes even more sensitive to $\Delta$.
A way around to compensate that would be to individually manipulate either 
$g_{\nu}$ or $\omega_{\nu}$ [cf. Eq. (\ref{weff})] though this would not work out
very efficiently. 
First, note that 
$\omega_{\nu}$ is also present in the denominator of Eq. (\ref{weff}).
Also, one must be careful when tuning $g_{s}$ and $g_{r}$ in order to maintain
the sender and receiver off-resonantly coupled to the channel.
Normally, the scale $\Delta$ imposes is such that 
it would become necessary to increase one of the outer 
couplings $g_{\nu}$ quite considerably thus disturbing a few normal modes in the neighborhood
of the $\omega_{\nu}$ level thereby breaking down the validity of the effective description in Eq. (\ref{Heff2}).
Besides all that, in principle there is no way to predict, sample by sample,
the specific disordered outcome 
so we are better off if we 
just fix $g_{\nu}$ and $\omega_{\nu}$ to some convenient value.
We also remark that when dealing with quantum communication protocols of this kind in 
spin chains \cite{bose03}, it is important to keep the level of external control over the system
to a minimum. Initialization and read-out procedures are the only forms
of control that should be allowed.  

All the things discussed above is valid for the case of standard uncorrelated stochastic fluctuations in the system`s parameters. However, a given disordered set of parameters might not be always
uncorrelated, say, site-independent \cite{dechiara05,burgarth05}. 
%
Let us now discuss some possible 
consequences of correlated disorder sequences on the channel, 
particularly those displaying
long-range correlations with power-law spectrum such as the ones generated from
Eq. (\ref{disorder}).
In this case, for 1D tight-binding models 
it is known that   
the underlying structure of the series induces the appearance of a set of delocalized
states around the middle of the band with well-defined mobility edges \cite{demoura98} provided $\alpha > 2$.
In order to elucidate that, we numerically calculate the normalized participation ratio distribution, for every eigenstate of Hamiltonian (\ref{Hchannel}), defined by
\begin{equation}
\xi_{k} = \dfrac{1}{N\sum_{i=1}^{N}|\langle i \vert E_{k} \rangle|^4},
\end{equation}
which assumes $1/N$ for fully-localized states and $1$ for uniformely extended states (that is
$\langle i \vert E_{k} \rangle = 1/\sqrt{N}$ $\forall$ $i$). 

%
\begin{figure}[t!] 
\includegraphics[width=0.45\textwidth]{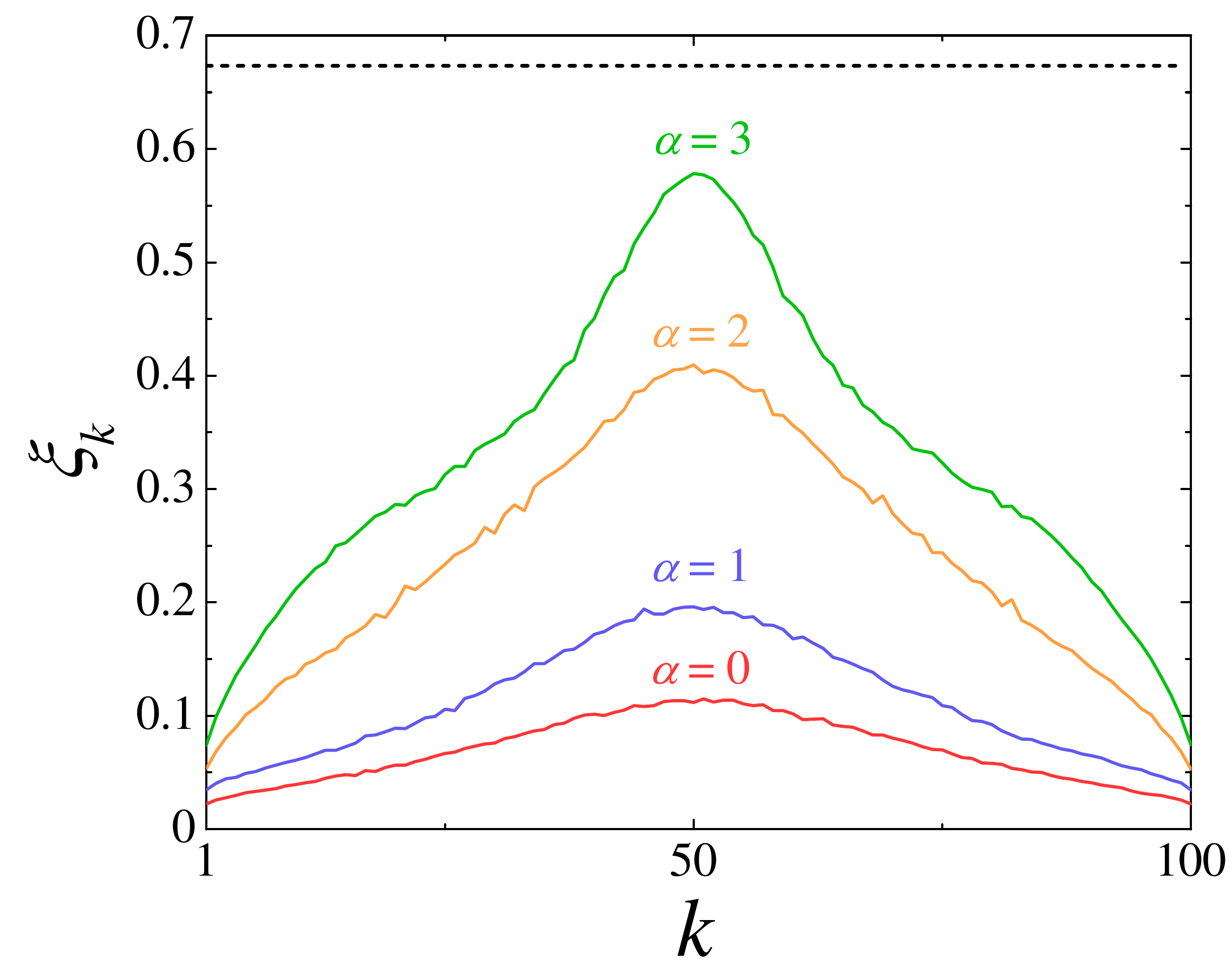}
\caption{\label{fig1} Normalized participation ratio of eigenstates $\xi_{k}$ for different values of $\alpha$ for a uniform channel ($J_{i,j} = J$)
composed of $N=100$ spins with on-site disorder given by Eq. (\ref{disorder}). Each curve was obtained by exact numerical diagonalization of Eq. (\ref{Hchannel})
and further
averaging $\xi_{k}$
over $10^{3}$ independent realizations of disorder. Note that  $\xi_{k}$ goes from $0.01$ for completely localized states to 
$1$ for uniformely extended states. 
For comparison, the dashed curve above (constant at $2/3$) shows the participation ratio distribution for the harmonic eigenstates of the noiseless case.
Here we clearly see that the presence of long-range correlations in the disorder distribution
promotes the appearance of delocalized states around the middle of the band
}
\end{figure}

Figure \ref{fig1} shows the resulting $\xi_{k}$ distribution (averaged over $10^{3}$ independent samples) as the degree of long-range 
correlations $\alpha$ is increased for a on-site-disordered channel consisting of $N = 100$ spins, including the noiseless case ($\omega_{n}\rightarrow 0$) for comparison (dashed line).
Note that we are considering the channel Hamiltonian only [cf. Eq. (\ref{Hchannel})], with $g_{\nu} = 0$.
Indeed, a prominent set of delocalized eigenstates
builds up around the band center. 
First of all, 
we should remark that the slight deflection of the $\alpha = 0$ curve (uncorrelated disorder) is solely due to the well known fact that the states at the band edges are more localized than those near the band center.
This gets much more pronounced when $\alpha = 2$ and higher, as expected.
Indeed, $\alpha > 2$ sets the transition point from an insulator to a metallic phase 
in eletronic tight-binding models, characterized by vanishing Lyapunov coefficients in the central part of the spectrum \cite{demoura98}. 
This happens exactly when 
the sequences generated by Eq. (\ref{disorder}) display persistent increments
according to the Hurst classification scheme \cite{fractalsbook}.

\begin{figure}[t!] 
\includegraphics[width=0.45\textwidth]{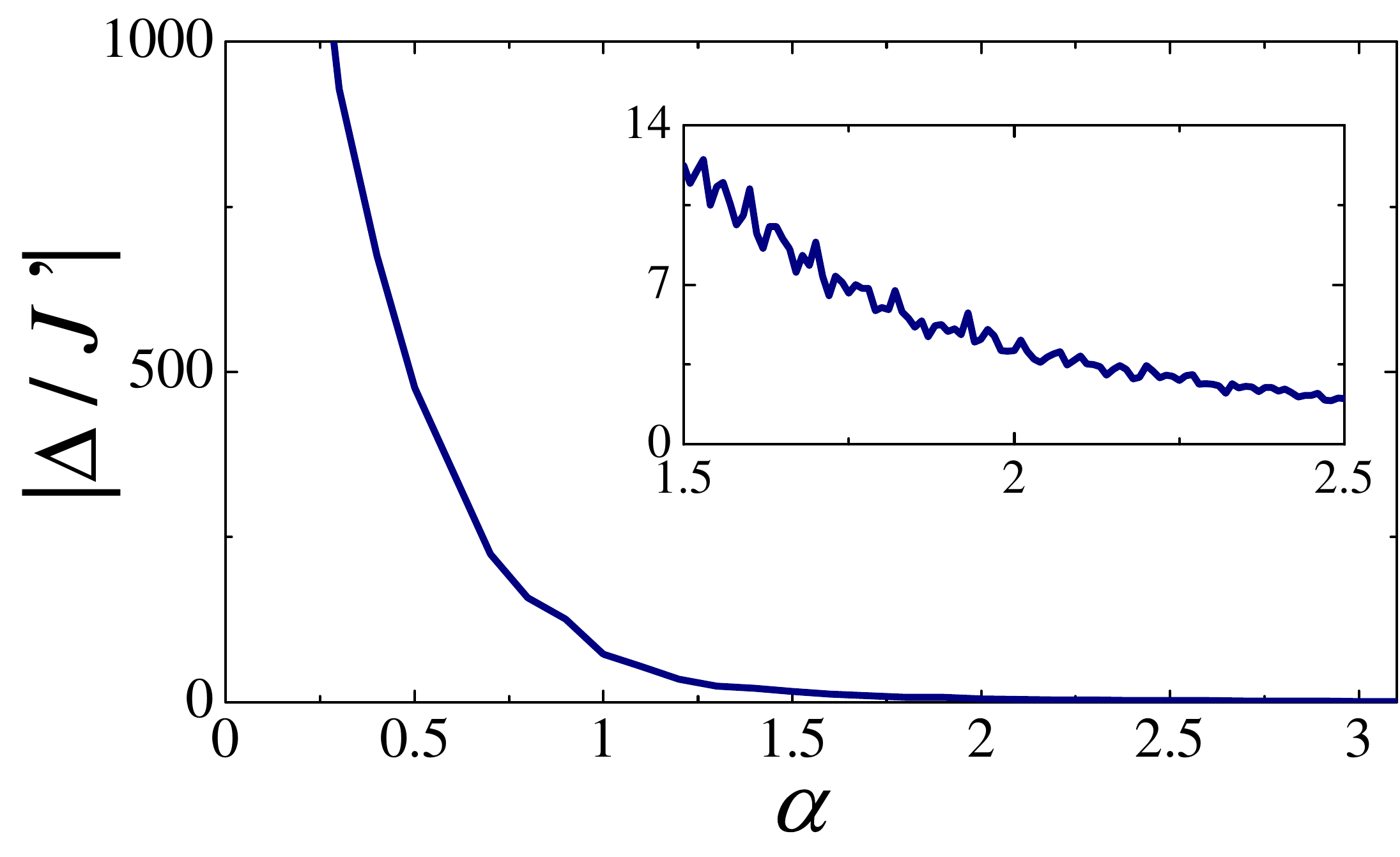}
\caption{\label{fig2} 
(Color online) Absolute value of detuning $\Delta = h_{s}-h_{r}$ in units of $J'$ versus $\alpha$ averaged over $500$ independent realizations of disorder for a channel featuring $N = 50$ spins with $\omega_{\nu} = 0$. 
Results were obtained from exact numerical 
diagonalization of Hamiltonian (\ref{Hchannel}).
Naturally, we are assuming very small
outer couplings $g_{\nu}$ so that the effective description, Eq. (\ref{Heff2}), holds.
The inset shows the very same quantity, zoomed in into the $\alpha =1.5$ to $\alpha=2.5$ interval.
}
\end{figure}

The likelihood of delocalized states in the presence of substantial amounts of 
disorder, not to mention the lack of mirror symmetry due to the on-site magnetic field distribution across the chain, sounds quite appealing. 
It means there is a suitable region in the frequency band of the 
channel -- in our case, in the middle of it, as seen in Fig. \ref{fig1} --
to tune the sender and receiver spins with. The corresponding eigenstates, featuring a delocalized nature, will display a broader amplitude distribution with greater
\textit{balance} between $a_{rk}$ and $a_{sk}$ thereby increasing the chances of inducing
a small detuning $\Delta$ [cf. Eqs. (\ref{Heff2}), (\ref{weff}), and (\ref{Jeff})], which is crucial for
having very high transfer fidelities. Figure \ref{fig2} shows how 
the absolute value of the ratio $\Delta/J'$
(averaged over several samples) behaves with $\alpha$ thus leaving no doubt the onset of long-range correlations
establishes a suitable ground for carrying out quantum communication protocols 
with weakly-coupled parties. 
As discussed earlier, uncorrelated fluctuations ($\alpha=0$) 
rules out any possibility of doing so, the ratio being extremely high. 
Things then get rapidly improved with $\alpha$ suggesting that already for $\alpha > 2$
one should obtain satisfying outcomes in the QST protocol, as we show in the following section.
%




%

\section{\label{sec5}Quantum-state transfer protocol}

The standard QST protocol goes as follows \cite{bose03}. Suppose that Alice is able to control the spin located at position $s$ and wants to send an arbitrary qubit
$\ket{\phi}_{s} = \alpha \ket{\downarrow}_{s}+ \beta \ket{\uparrow}_{s}$ to Bob which has
access to spin $r$. Now let us assume that the rest of the chain is initialized
in the fully polarized spin-down state so that 
the whole state reads
$\ket{\Psi (0)} = \ket{\phi}_{s}\ket{\downarrow}_{1}\ldots \ket{\downarrow}_{N}\ket{\downarrow}_{r}$. She then let the system evolve 
following its natural dynamics, 
$\ket{\psi(t)} =\hat{\mathcal{U}}(t)\ket{\psi(0)}$, 
where $\hat{\mathcal{U}}(t)\equiv e^{-i\hat{H}t}$ is the unitary time-evolution operator.
Ideally, she expects that at some prescribed time $\tau$ 
the evolved state takes the form
$\ket{\Psi (\tau)} = \ket{\downarrow}_{s}\ket{\downarrow}_{1}\ldots \ket{\downarrow}_{N}\ket{\phi}_{r}$. At this point, 
Bob receives state $\rho_{r}(\tau) = \mathrm{Tr}_{s,1,\ldots,N}\ket{\Psi (\tau)}\bra{\Psi (\tau)}$
and thus the transfer fidelity can be evaluated by $F_{\phi}(\tau) = \bra{\phi} \rho_{r}(\tau) \ket{\phi}$
Note, however, that this measures the performance of QST for a specific input. 
In order to properly evaluate the efficiency of the \textit{channel}, we may average the above quantity over all input states $\ket{\phi}_{s}$ (that is, over the Bloch sphere)
which results in \cite{bose03}
\begin{equation}\label{avF}
F(t) = \dfrac{1}{2}+\dfrac{f_{r}(t)}{3}+\dfrac{f_{r}(t)^2}{6}
\end{equation}
for an arbitrary time with $f_{i}(t) \equiv \vert \bra{i} e^{-i\hat{H}t} \ket{s} \vert$
%
Therefore, we note that such a state-independent figure of merit of QST depends solely upon the transition amplitude between the sender and receiver spins with $F(t)=1$ only when
$f_{r}(t) = 1$. 
The problem of transmitting a qubit state from one point to another can thus be viewed as 
a single-particle continuous quantum walk \cite{kemperev} on a network and the goal is to 
find out ways to transfer the excitation between two distant nodes
with the highest possible transition amplitude. 

In the case of weakly-coupled spins in which an effective two-site 
interaction sets in [cf. Eq. (\ref{Heff2})], 
the transition amplitude $f_{r}(t)$ will strongly depend upon the resonance between
$h_{s}$ and $h_{r}$, that is $\Delta$. In the previous section, we have seen
that the emergence of long-range correlations (see Fig. \ref{fig2}) 
favors smaller values of $\Delta$. 
Now, let us finally see about the resulting QST performance.
As a testbed, we consider a $N=50$ channel, $g_{\nu}=g= 0.001$ (in units of $J$),
and $\omega_{\nu} = 0$. 
Given the size of the channel, this chosen value for $g$ assures that
the subspace created by states $\ket{s}$ and $\ket{r}$ becomes safely shielded from influence of 
channel normal modes lying around the band center. Even if one of them gets close by, 
it is very likely that the eigenstate will not be extremely asymmetric 
due to the presence of delocalized states
for high enough $\alpha$ [see Fig. \ref{fig1}].  

%
\begin{figure}[t!] 
\includegraphics[width=0.45\textwidth]{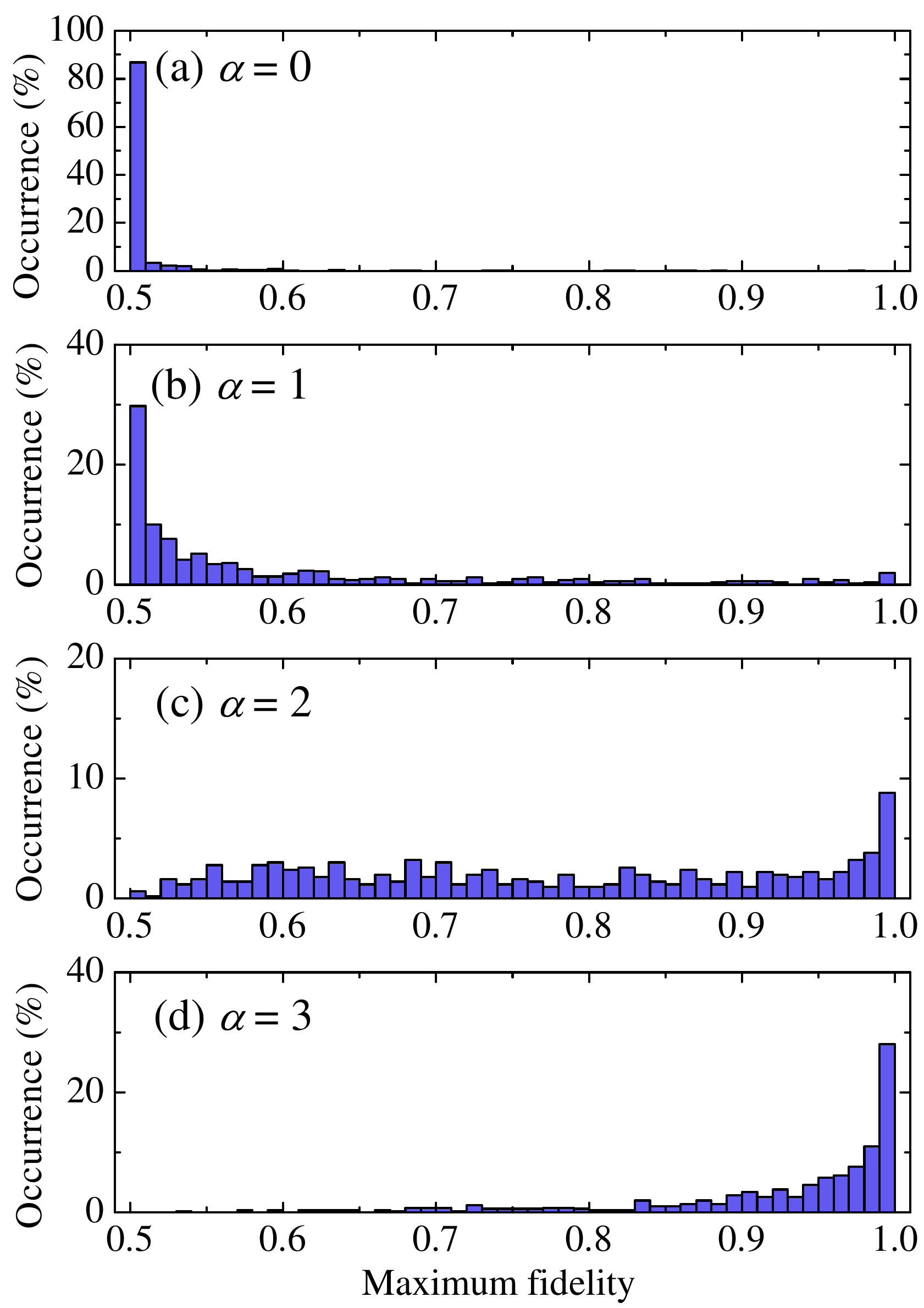}
\caption{\label{fig3} 
(Color online) Maximum-fidelity histogram for 500 indepedent realizations of disorder for $\alpha=0,1,2,$ and $3$. 
Results were obtained from exact numerical diagonalization of the full Hamiltonian $\hat{H} = \hat{H}_{\mathrm{ch}}+\hat{H}_{\mathrm{int}}$
with $N=50$, $\omega_{n}$, $n=1,\ldots,N$, given by Eq. (\ref{disorder}), $\omega_{\nu} =  0$, $g_{\nu} = g = 0.001$ (in units of $J$). 
The maximum fidelity $F_{\mathrm{max}} = \mathrm{max}\lbrace F(t)\rbrace$ [see Eq. (\ref{avF})] was registered during time interval
$[0,20\tau]$, with $\tau = \pi/(2g^{2})$.
}
\end{figure}

In Fig. \ref{fig3} we show the
sample distribution of the maximum fidelity $F_{\mathrm{max}} = \mathrm{max}\lbrace F(t)\rbrace$
[as defined above in Eq. (\ref{avF})] 
achieved in time interval $t \in [0,20\tau]$, 
with $\tau=\pi/(2g^{2})$ being the corresponding time (in units of $1/J$) for which
a complete transfer would occur for the noiseless case, $f_{r}(\tau) \approx 1$,
as seen in Sec. \ref{sec3}. 
That interval is a pretty reasonable one in order to 
guarantee at least one full Rabi cycle in most of the samples. 
Recall that the effective sender-receiver hopping strength $J'$ dictates the time scale of the dynamics
and is strongly affected by disorder.
Figure \ref{fig3} ultimately confirms what it has been suggested 
by Fig. \ref{fig2}. Indeed, strong long-range correlations 
in the disorder distribution enhances the 
figure of merit of QST enormously.
Even more impressive is the fact that, for $\alpha = 2$ and $\alpha=3$ [see Figs. \ref{fig3}(c) and \ref{fig3}(d), respectively],
we find the number of occurrences of fidelities $F_{\mathrm{max}}\approx 1$ to be the highest one. 
We also note that the fidelities for $\alpha=2$ case [Fig. \ref{fig3}(c)] is fairly well distributed 
across all the possible outcomes, thus indicating 
a transition regime.

%
\begin{figure}[t!] 
\includegraphics[width=0.35\textwidth]{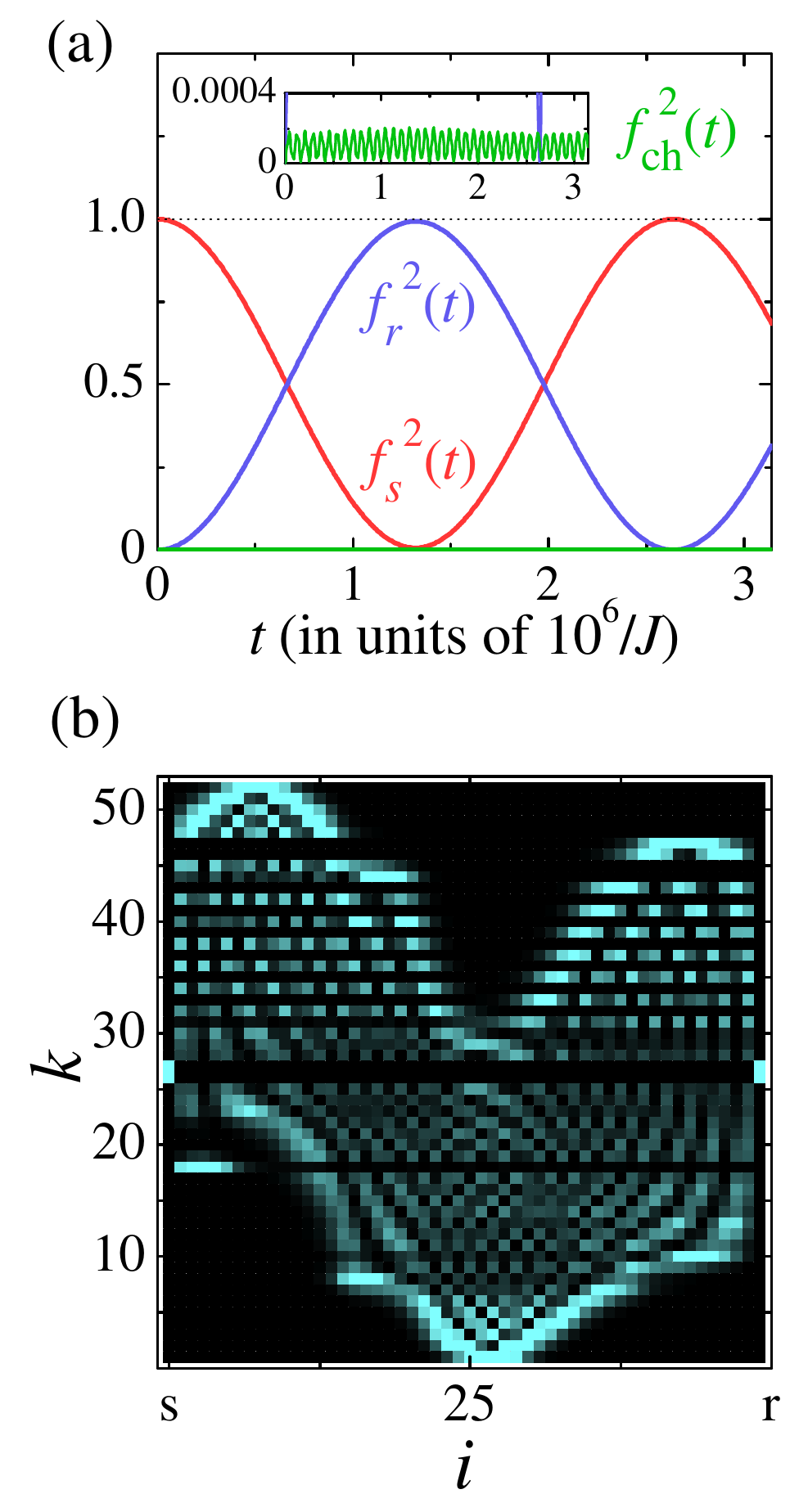}
\caption{\label{fig4} 
(Color online) (a) Time evolution of the occupation probability amplitudes for the sender $f_{s}^{2}(t)$, receiver $f_{r}^{2}(t)$, and channel spins, the latter being
the sum of the amplitudes within the channel,
namely $f_{\mathrm{ch}}^{2}(t)\equiv \sum_{n=1}^{N} f_{n}^{2}(t)$. 
For this particular realization, we simply took one of 
the samples which provided $F_{\mathrm{max}}\approx 1$ in Fig. \ref{fig3}(d) for $\alpha=3$. Note that the time scale has been reduced 
to twice the transfer time for the noiseless case, that is $2\tau$,
for a better view of a Rabi-like cycle. The inset shows the very same graph but for 
a much smaller scale of amplitude in order to account for the
(rather negligible) behavior of 
$f_{\mathrm{ch}}^{2}(t)$.
(b) Corresponding density plot of
the eigenstate spatial distribution
$\vert \langle\vert i \vert \psi_{k} \rangle \vert^{2}$ for every $k$ (in increasing order of energy). Darker (brighter) spots indicate lower (higher) overlaps. 
Note the formation of a pair of states in the middle of the band with strong overlap in
$\ket{s}$ and $\ket{r}$ simultaneously. These are the source of 
such high-fidelity QST rounds.
}
\end{figure}

In order to provide an explicit view on what is actually going on in the QST process, in Fig. \ref{fig4}(a) we show the time evolution of the 
occupation probabilities $f_{i}^{2}(t)$ of the sender ($i=s$), receiver ($i=r$), 
and channel [$f_{\mathrm{ch}}^{2}(t)\equiv \sum_{n=1}^{N} f_{n}^{2}(t) $] spins
for one
particular (ordinary) sample, out of many successful ones
(meaning $F_{\mathrm{max}}\approx 1$) encountered for $\alpha=3$ [see Fig. \ref{fig3}(d)].
There we see a genuine Rabi-like behavior yielding a very high-quality QST.
We reduced the time scale to $2\tau$ 
so we can
have a more detailed view on a complete cycle. 
Therefore, in this case the transfer time happens to be 
roughly the same as for the noiseless case.
Further, we note that the channel is barely populated for all practical purposes [see the inset of Fig \ref{fig4}(a)], 
meaning that Eq. (\ref{Heff2})
is a robust approximation. Those residual beatings seen 
for $f_{\mathrm{ch}}^{2}(t)$
in are due to some negligible mixing between both channel
and sender/receiver subspaces. 
One could get rid of it by further decreasing $g$. 
Care must taken, though, not to compromise the transfer time scale since 
it increases $\propto\ 1/g^{2}$. 

Figure \ref{fig4}(b) shows the corresponding spatial distribution of eigenstates,
$\vert \langle\vert i \vert \psi_{k} \rangle \vert^{2}$, along the whole spectrum $k$.   
First, note that the outer parts of the spectrum are
mostly populated by localized-like eigenstates. Indeed, the eigenstates get more delocalized
as we move towards the center of the band, as discussed before [see Fig. \ref{fig1}].
We also point out the asymmetrical aspect of the eigenstate distribution. 
Still, it turns out to be possible to span an independent subspace involving only the sender and receiver
spins [Eq. \ref{Heff2}] so that their corresponding eigenstates become 
close to $\ket{s}\pm \ket{r})/\sqrt{2}$. 
By looking closely at Fig \ref{fig4}(b), we also spot a few eigenstates showing strong asymmetries between  
spins $1$ and $N$.
Fortunately, since $a_{sk}$ and $a_{rk}$ is fairly balanced across the spectrum and
due to the fact that
the channel eigenstates lying around the middle of the band (less asymmetric) have great
influence on $\Delta/J'$, given that 
the terms in the sum in Eqs. (\ref{weff}) and (\ref{Jeff}) goes $\sim 1/E_{k}$,
the sender and receiver spins are able to find a way out through such 
asymmetries and establish an effective resonant interaction between them
thus resulting in an almost perfect QST for most of the samples.   

%
\begin{figure}[t!] 
\includegraphics[width=0.45\textwidth]{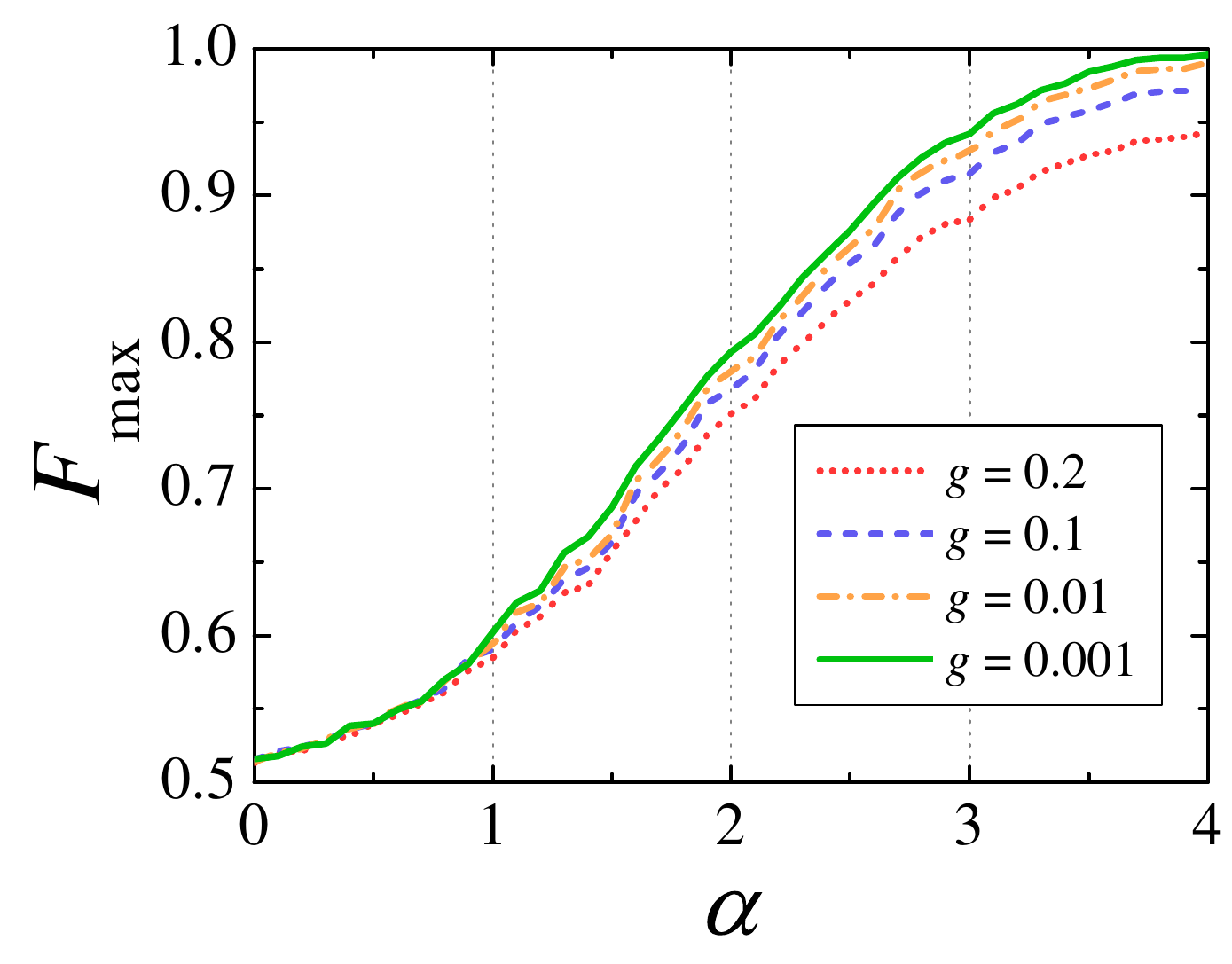}
\caption{\label{fig5} 
(Color online) Maximum fidelity versus $\alpha$ averaged over 500 independent disorder realizations. Now, we have set $g = 0.001$, $0.01$, $0.1$, and $0.2$ (in units of $J$)
while other system's parameters were kept as usual, namely
$N=50$ and $\omega_{\nu} =  0$. 
The maximum fidelity $F_{\mathrm{max}} = \mathrm{max}\lbrace F(t)\rbrace$ for each sample was again obtained during time interval $[0,20\tau]$, with $\tau = \pi/(2g^{2})$.
}
\end{figure}

Last, in order to evaluate 
a representative outcome for $F_{\mathrm{max}}$ for a given $\alpha$,
in Fig \ref{fig5} we plot its average 
over all the samples 
for a large window of $\alpha$ values. 
This clearly illustrates  
the overall behavior of the occurrences of $F_{\mathrm{max}}$ 
as one increases the degree of long-range correlations in the disorder distribution. 
Note that we are also showing the curve for many values of $g$, only to stress the importance of setting this parameter
as smaller as possible
so as to avoid mixing between the channel and sender/receiver subspaces.  
Indeed, we see quality of QST is affected by that. 
As we go towards smaller values of $g$, 
there is a saturation point indicating that
Hamiltonian \ref{Heff2} has reached its final form. It means that if we keep on decreasing $g$, the QST fidelity will not get any better and the time scale of the transfer will
increase substantially.
%
Finally, we identify in Fig. \ref{fig5}
that the $F_{\mathrm{max}}$ growth profile 
is more pronounced
between $\alpha = 1$ and $\alpha = 3$
until it saturates for higher values of $\alpha$. 
This is associated to the fact that
the long-range correlated sequence generate by Eq. (\ref{disorder}) becomes 
nonstationary for $\alpha > 1$ and acquires persistent
character when $\alpha>2$, thereby triggering 
the appearance of delocalized states in the middle 
of the band \cite{demoura98, adame03}. 
  
\section{\label{sec6}Concluding remarks}

We studied a QST protocol through a 
$XX$ spin channel with on-site long-range-correlated disorder. 
The protocol involved a couple of communicating spins weakly coupled to the channel not
matching with any of its normal modes so that the transfer takes 
place through Rabi-like oscillations between
the ends of the chain \cite{wojcik05, almeida16}.
We focused on the reduced sender/receiver description based on Hamiltonian (\ref{Heff2})
which embodies all the relevant information regarding the way 
they are affected by the channel, thus allowing one to foresee
the QST outcome based on the renormalized parameters contained in the two-site effective Hamiltonian.
   
We showed that this class of 
weakly-coupled models are indeed robust against external perturbations \cite{yao11} as the effective interaction between sender and receiver spins do not depend upon
the entire wavefunction of the spectrum but rather on the local amplitudes of the spins they are connected to. 
Because of that, 
we realize we do not necessarily need a perfect symmetric chain to 
to achieve an almost perfect QST.
When scale-free correlations with a power-law spectral density $S(k)\propto k^{-\alpha}$ set in, 
the disorder distribution is such that it can support delocalized eigenstates around
the center of the band \cite{demoura98}. 
Those are able to provide a broader, more balanced distribution
of amplitudes even in the presence of asymmetries, what makes it possible
to induce effective resonant interactions between $\ket{r}$ and $\ket{s}$,
provided $\alpha$ is high enough, thus resulting in extremely high 
fidelities, with most of the samples providing $F_{\mathrm{max}} \approx 1$. 

Note that we have not considered the case of structural disorder here, that is, 
fluctuations on the spin couplings. However, on-site disorder actually
embodies a worst-case, and hence more realistic, scenario since the spectrum also looses 
its symmetry, differently from structural fluctuations.

We remark that disorder, either correlated or not, might arise naturally due to
experimental imperfections in the
manufacturing process of solid state devices for quantum information processing. 
However, we may also think about inducing those correlations
somehow since, as we have shown, 
it may not be so detrimental for certain communication tasks
as in the uncorrelated-disorder scenario.
Overall, it should be easier to allow for that than designing a chain
with a very specific set of parameters, which demands a high degree of control.  
Our work further promotes the study of quantum communication protocols
in disordered, asymmetric, spin chains. 

%

\section{Acknowledgments}

This work was partially supported by CNPq (Grant No. 152722/2016-5),
CAPES,  FINEP,  and FAPEAL (Brazilian agencies).


%

\end{document}